# Excitation of surface plasmon polaritons guided mode by Rhodamine B molecules doped in PMMA stripe

**D.G. Zhang [1,3], X.-C.Yuan[2], A. Bouhelier[4], P.Wang[3] and H.Ming[3]**

*1School of Electrical and Electronic Engineering, Nanyang Technological University, Nanyang Avenue, Singapore 639798*

*2Institute of Modern Optics, Key Laboratory of Optoelectronic Information Science & Technology, Ministry of Education of China, Nankai University, Tianjin 300071, People's Republic of China*

*3Institute of Photonics, University of Science and Technology of China, Hefei, Anhui, 230026, People's Republic of China*

*4Institut Carnot de Bourgogne, CNRS-UMR 5209, Université de Bourgogne, 21078 Dijon, France*

*[*]Corresponding author: xcyuan@nankai.edu.cn*

In this letter we show the inclusion of Rhodamine B molecules (RhB) inside a dielectric-loaded surface plasmon waveguide enables for a precise determination of its optical characteristics. The principle relies on the coupling of the fluorescence emission of the dye to plasmonic waveguided modes allowed in of the structure. Using leakage radiation microscopy in real and reciprocal spaces, we measure the propagation constant of the mode and as well as their attenuation length.

*OCIS codes: 240.6680, 250.5460, 260.2510.*

Surface plasmon polaritons (SPPs), originating from a resonant interaction between surface charge oscillations and an electromagnetic field, are opening up new directions in the field of optoelectronics [1]. In this context, SPPs waveguides are one of the key elements enabling transfer of information across a surface[2, 3, 4]. Recently, the concept of dielectric–loaded surface plasmon polariton waveguide (DLSPPW) was introduced to increase the lateral confinement of surface plasmon in waveguiding structures. A DLSPPW typically consists of a dielectric stripe deposited on a metal film [4, 5, 6, 7]. Common dielectric materials used in DLSPPWs are SiO2 or PMMA that provide a reasonable tradeoff between field confinement and propagation distance. Efficient integration of passive components based on this technology was recently demonstrated[3, 8, 9, 10]. The main advantage of DLSPPW over other type of plasmonics waveguides resides in the fact that the dielectric layer can be artificially doped by suitable materials so that the optical properties of the waveguide could be externally modified. This provides a mean to actively control[2, 11] a SPP mode or even partially compensating losses by stimulated emission of surface plasmon[12]. In this letter, a PMMA film doped with RhB molecules was patterned by electron beam lithography to form a PMMA stripe with a rectangular section on the silver film. Since fluorescent molecules placed near a metallic surface can readily transfer their radiation to surface plasmon, the RhB included into PMMA stripe works as an excitation source for SPPs supported at the interfaces of Ag/air or air/PMMA/Ag.

The samples were prepared as follow. RhB molecules (0.1mg/ml) were dissolved in a PMMA solution (950K PMMA, Solids: 2% in Anisole, from MICRO.CHEM corp.) for about 48 hours. The doped solution was agitated by ultrasonic disrupter for 30 minutes before the spin-coating process. The RhB doped PMMA film used in this experiment was obtained by spin coating the solution onto 45 nm-thick silver films deposited by electron beam evaporation on a glass substrate. The film was baked for 10min at 105°C to remove the solvent. Electron beam lithography (Raith GmbH, e_LiNe) and standard developing procedure were used to write PMMA stripes.

A 532 nm wavelength laser was used to excite the RhB molecules by using an oil-immersion objective (60X, N.A., 1.42). The laser beam was expanded with a lens assembly to overfill the back aperture of the objective. Excitation of the RhB and detection of the fluorescence signal was collected by the same objective (Epi-configuration). The polarization of the excitation was controlled by a combination of a half-wave plate and a polarizer. The power of the 532nm laser before entering the objective was about 0.1mW. A 532 nm long pass edge filter was placed before a charge-coupled device (CCD) camera to reject the exciting beam. Both the Fourier plane image and direct image plane were captured by the CCD cameras as shown in Fig. 1. An atomic force image of the fabricated stripe is shown in inset of the Fig.1. The width, thickness and length of the dielectric stripe are 160nm, 68nm and 60 μm respectively.

Figure 2 (a) shows the fluorescence image of RhB doped PMMA stripe excited by the tightly focused 532nm beam with vertical polarization. We recall here that the image represents the light emitted through the silver film inside the glass substrate (leakage radiation). The white dash-line circle represents the position of the beam focus positioned at a location along the PMMA stripe. Figure 2 (a) shows that a waveguided mode exists inside the stripe and propagates on either side of the focal spot. Figure 2 (b) is the intensity profile along the strip corresponding to the white dash line in Figure 2 (a). The attenuation length is obtained by fitting the profile with a single exponential function. The intensity of the waveguided modes vanishes with distance with an exponential decay of 6.4 μm typical from surface plasmon mode.

To verify that the signal detected at the RhB signal originates from SPPs wave, an image of the Fourier plane of the microscope was acquired to determine the wave-vector content of the emission (Figure 2(c)). The emission is distributed in precise location of the reciprocal space. Part of the emission wave-vectors is emitted in an almost complete bright ring. This ring represents the commonly known surface plasmons coupled emission (SPCE) [13]. Based on the know N.A. of the objective, the wave-vector (K) of the emitting fluorescence inside the ring can be estimated as $1.036K_0$ ($K_0$ is the vacuum wave-vector of the RhB fluorescence peak taken at 576 nm). This value is approximately equal to the calculated KSP (1.044 $K_0$) of SPPs (n= 0.12+3.547i for Ag) that would be excited at Ag/Air interface at 576 nm. However, it should be noted that the ring of SPCE is caused by molecules that are inside the PMMA stripe. To verify that no molecules or thin PMMA layer was residual on the silver after the development process, we moved the focus point out of the PMMA strip and do not observe noticeable emission in the Fourier-plane image. We conclude that the RB molecules located inside the waveguide can excite SPPs on the adjacent Ag/Air interfaces. The SPCE wave-vectors are distributed on the ring because of the random orientation of the molecules inside the waveguide. Note that the ring is not fully complete at the radii along the vertical axis. Here the emission wave-vectors are distributed into two diametrically-opposite lines. These lines are the signature of guided $TM_{00}$ modes[2] with a constant effective index propagating upwards and downwards from the focal spot (arrows in Fig. 2(a)). The wave-vector of this SPP mode is evaluated at 1.103 $K_0$. The full-width at half-maximum of the line (0.092 $K_0$) does not accurately represent the losses of the mode since the emission is distributed among the spectral width of the emitting fluorescence. We can conclude from Fig. 2 that the fluorescence is coupled to SPCE and also to SPPs mode propagating inside the PMMA stripe. Due to the finite thickness of the silver layer, the modes are leaking in the substrate enabling thus imaging of their intensity.

We have investigated another case, where the excited beam was focused onto the edge of a doped PMMA extended film. Figure 3 (a) shows the direct-space leakage radiation image. The left part of the image corresponds to the RhB doped PMMA film (about 80nm thickness) on silver film while the right part is the bare silver. The center white line marks the edge of PMMA layer. Figure 3 (a) shows that there is an attenuated wave propagating from the edge to the right part. Figure 3 (b) is the corresponding image of the Fourier plane showing a bright full ring and a right-hand side of a ring. The wave-vector of the half ring is measured at 1.03 $K_0$. This half-ring can be attributed to the SPCE on the interface of air/Ag. The wave-vector of the full ring is measured at 1.419 $K_0$, which can be attributed to the SPCE on the interface of Air/PMMA/Ag.

This phenomenon is different from that shown in Figure 2(c), where two SPPs waves propagated on both side and the entire ring appear due to the two air/Ag interface on both sides of the PMMA strip. Figure 3 also clearly gives out the physics picture of the SPCE process which can not be observed with prism based set-up[14]: At the interface of PMMA and air marked with the white line, excited RB molecules at the side of the PMMA film transfer to the SPPs on the Air/Ag interface, during the propagation of the SPPs wave, it leaky radiates to the far field at the same angle of the SPPs resonance angle, which is so called the directional emission of the fluorescence coupled with SPPs (SPCE).

Relying on the above observations, we fabricated a two-way coupler constituted of two orthogonally aligned 30μm-long PMMA stripes as shown in the bright field transmission image of Figure 4(a). The width and thickness of the stripes are 200nm and 70nm respectively. The exciting laser beam was focused onto the intersection point, so SPPs mode can be excited simultaneously in the two orthogonal stripes. Figure 4(b), (c) and (d) are the direct-space fluorescence images acquired for three different polarization direction of the exciting beam. The distribution of intensity inside the two stripes depends on the orientation of the incoming polarization: maximum coupling in a given waveguide is obtained for a collinear polarization. We defined a concept of intensity ratio as following to qualitatively describe the phenomenon: the intensity ratio of two points at the two arms with equal distance to the intersection point (5 m in this experiment). The measured intensity ratio (horizontal arm vs. vertical arm) in Figure 4(a), (b) and (c) are about 4.3, 1.0 and 2.7 respectively. Our observation provides a mean to modify the intensity ratio inside two-way couplers simply by changing the polarization direction of the exciting beam.

In summary, RhB-doped PMMA stripes on silver films were successfully fabricated and their optical properties were characterized with leakage radiation microscopy. Experimental results show that the available modes sustained by the system are excited by molecules and can serve as basis for rapid quantitative determination of the properties of the plasmonic waveguides. The gain materials doped DLSPPW also has potential application in SPPs related waveguide amplifier and waveguide laser, which are key elements in integrated photonics[15,16].

This work was partially supported by the National Natural Science Foundation of China under Grant No. (60778045), the National Research Foundation of Singapore under Grant No. NRF-G-CRP 2007-01 and the Ministry of Science and Technology of China under Grant no. 2009DFA52300 for China-Singapore collaborations. XCY acknowledges the support given by Nankai University (China) and Nanyang Technological University (Singapore). HM acknowledges the funding support given by the National Natural Science Foundation of China under Grant No.60736037 and the National Key Basic Research Program of China under Grant No. 2006CB302905.

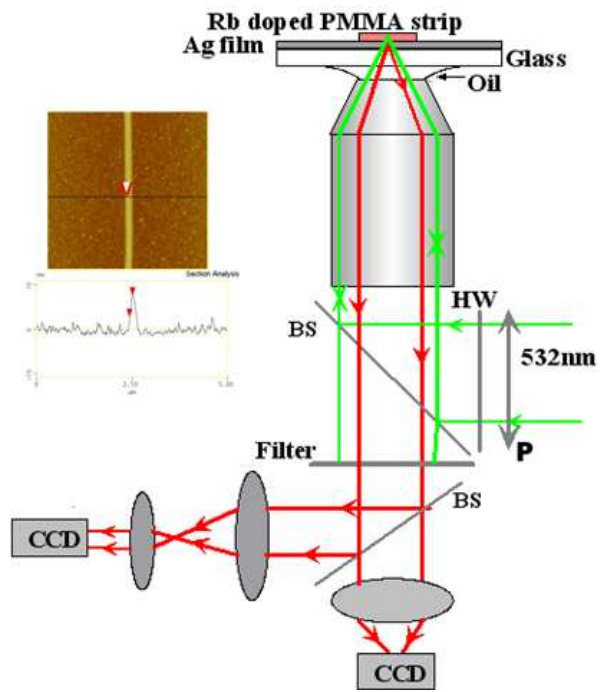

FIG.1. Sketch of the experimental set-up, BS is beam-splitter. HW is half wave-plate, P is polarizer. Inset graph is the AFM image of the stripe on the silver film.

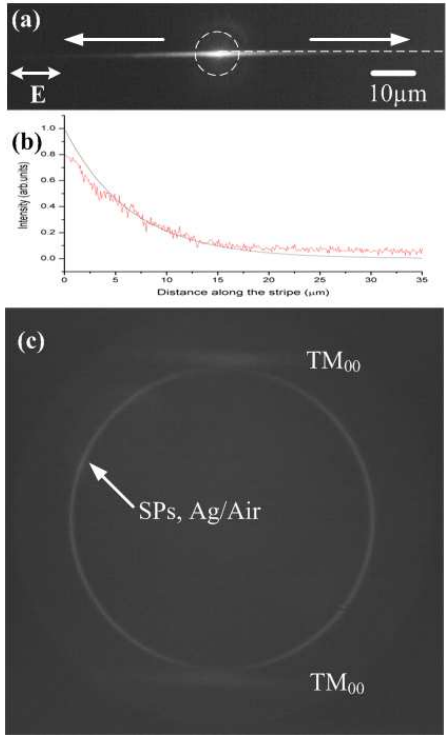

FIG.2. (a) Leakage radiation micrograph of the fluorescent signal coupled in the waveguide. The white circle shows the location of the excitation spot. The two vertical arrows represent the direction of propagation of the two modes propagating away from the excitation spot. (b): Leakage radiation image of the Fourier plane of the microscope. The bright circle represents wave-vectors provided by the plasmon propagating at the Ag/air interface whereas the two horizontal lines are signature of guided $TM_{00}$ modes. The $n_{eff}$ of the SPPs at the Ag/Air and the $TM_{00}$ modes are about 1.036 and 1.103 respectively.

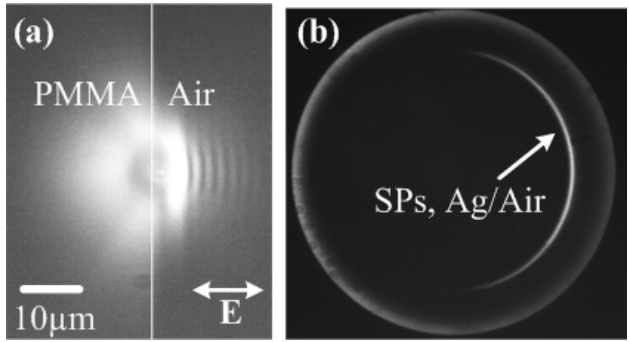

FIG. 3. Excitation of SPPs (Air/Ag and Air/PMMA/Ag interface) with RB doped in PMMA films (a): Florescence image, (b): Fourier plane fluorescence image.

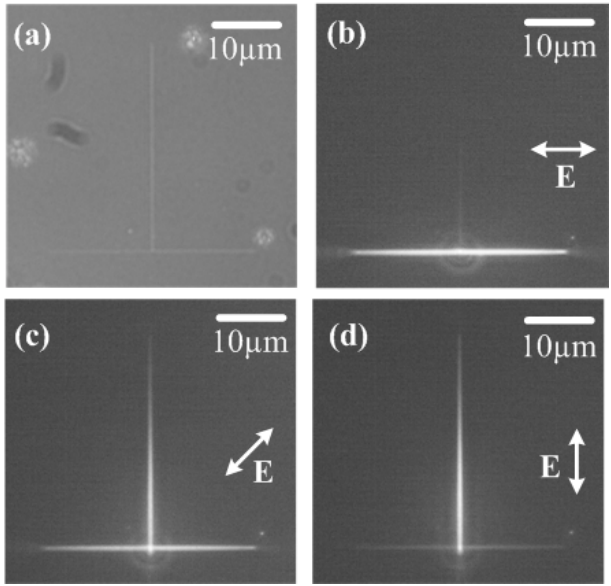

FIG.4. (a) Bright-field transmission image of two-way coupler. (b), (c) and (d) are direct-space fluorescence image for three different polarization of the exciting beam ($0^o$, $45^o$, and $90^o$ relative to the horizontal axis).